\documentclass[usenatbib,usegraphicx]{mn2e}

\title[L3CCDs and visibility parameter estimation]{Low light level
CCDs and visibility parameter estimation}

\author[A. Basden et al.]{A. G. ~Basden$^1$\thanks{E-mail:
abasden@mrao.cam.ac.uk} and C. A. ~Haniff$^1$\\
$^1$Astrophysics Group, Cavendish Laboratory, Madingley Road,
Cambridge CB3 0HE\\}

\begin{document}
\date{Released 2003 Xxxxx XX}

\pagerange{\pageref{firstpage}--\pageref{lastpage}} \pubyear{2003}
\label{firstpage}
\maketitle

\begin{abstract}
Recently, low light level charge coupled devices (L3CCDs) capable of
on-chip gain have been developed, leading to sub-electron effective
readout noise, allowing for the detection of single photon events.
Optical interferometry usually requires the detection of faint signals
at high speed and so L3CCDs are an obvious choice for these
applications.  Here we analyse the effect that using an L3CCD has on
visibility parameter estimation (amplitude and triple product phase),
including situations where the L3CCD raw output is processed in an
attempt to reduce the effect of stochastic multiplication noise
introduced by the on-chip gain process.  We establish that under most
conditions, fringe parameters are estimated accurately, whilst at low
light levels, a bias correction which we determine here, may need to be
applied to the estimate of fringe visibility amplitude.  These results
show that L3CCDs are potentially excellent detectors for astronomical
interferometry at optical wavelengths.
\end{abstract}

\section{Introduction}
Optical interferometry requires the detection of an interference
fringe pattern generated when starlight collected by separated
apertures is recombined with a spatially or temporally modulated
optical path difference (OPD).  If the total distance travelled by the
two beams of light is equal, or the OPD is a whole number of
wavelengths, constructive interference will result in an intensity
maxima.  Likewise, if the OPD is an odd number of half wavelengths,
destructive interference results in an intensity minima.  In a typical
interferometer, intensities are measured for a range of OPDs, with the
OPD differing by a distance of several wavelengths about the zero OPD.
Spatial and temporal modulation of the OPD can both be used by optical
interferometers, for example the Amber instrument on the VLTI (spatial
modulation, \citet{amber}) and the COAST (temporal modulation,
\citet{coast}).  An oscillating signal or fringe pattern results, the
contrast and phase of these fringes providing information about the
source, including the source diameter and shape, and, when the light
is combined using three or more apertures, information about asymmetry
within the object.  The entire fringe must be sampled within a few
times the time-scale of atmospheric perturbations (typically 10 ms in
the optical regime), resulting in low signal intensities, and for all
but the brightest stars the signal is very faint.  To date, the
faintest reported optical measurement using a separate element
interferometer is an object with visible magnitude of about 8.5 (I
magnitude 6.1) \citep{haniff}.  Increasing the sensitivity of
interferometers is therefore essential if they are to be used to
observe a wide range of scientific sources.  One way of doing this is
to observe over a wide range of wavelengths, allowing a greater light
throughput.  However, to maintain a useful coherence length, the
bandwidth of light on any detector element must be kept small,
requiring independent detection of many different wavelengths
(spectroscopic detection).  This will require at least a one
dimensional array of detector elements, which may be expensive and
fragile if conventional photon counting detectors (APDs and PMTs) are
used.

CCDs ought to be ideal detectors for spectroscopic interferometry
because they are stable devices available in large-format arrays,
allowing one CCD to replace many single element detectors.  However,
their major shortcoming is readout noise, i.e.\ the additional signal
added at the on-chip output amplifier where the photo-electrons are
counted.  A typical interferometer can require CCD readout rates of up
to and above 1 MHz since there will be many spectral channels, and
each of these must be sampled many times within an atmospheric
coherence time.  Currently, the best noise level achieved using
standard CCDs at these readout rates is typically 10 $\mathrm{e}^-$
\citep{jerram}.  Signal levels lower than this are then swamped by
noise.

Low light level CCDs (L3CCDs) provide a solution to this problem, with
an on-chip gain amplifying the signal prior to the on-chip readout
amplifier \citep{jerram}, resulting in a sub-electron effective
readout noise. This, combined with the high quantum efficiency (QE) of
L3CCDs (up to 90 percent), which compares favourably with the QE of
APDs \citep{takeuchi}, makes them prime candidates as interferometric
detectors.  Being array detectors they can be used for spectroscopic
detection allowing greater use of the available light, but they do
have the disadvantage of introducing an additional source of noise due
to the stochastic multiplication process. However for one application,
we have shown \citep{basden} that the magnitude of this noise can be
reduced if the L3CCD output signal is treated in the correct way, by
thresholding.

In this paper we investigate the use of L3CCDs for interferometric
fringe detection.  In most situations we find that no bias correction
is required to estimate fringe parameters correctly.  In some cases, a
correction is necessary when estimating visibility amplitude but we
provide this correction here.  We also develop a suitable treatment
allowing unbiased estimation of bispectrum (closure) phase.

The layout of the paper is as follows. In section 2 we present a
background to interferometric fringes and the use of L3CCDs, as well
as our modelling techniques. In section 3 we present results and our
conclusions are summarized in section 4.

\section{Interference and L3CCDs}
\subsection{Introduction to interferometric fringe production}
\begin{figure}
\includegraphics[width=8.5cm]{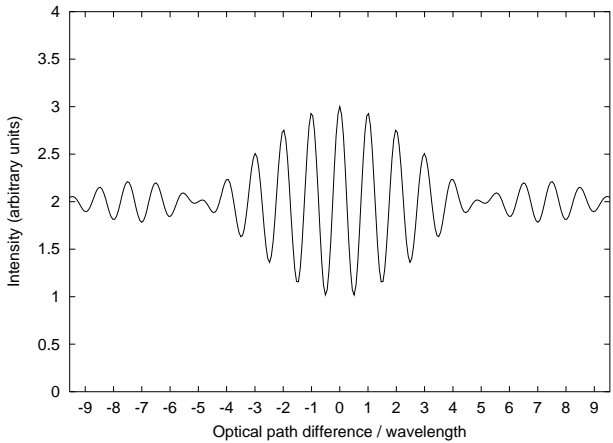}
\caption{Example of a simplified fringe pattern with a maximum
visibility amplitude of 0.5.  The enveloping shape of this fringe
pattern is due to a finite signal bandwidth, $\Delta
\lambda/\lambda=0.2$ in this case.}
\label{fringepic}
\end{figure}

In interferometry, the raw image we obtain is an interferometric
fringe pattern, which will show a periodic intensity variation due to
either spatial or temporal OPD modulation depending on the type of
interferometer.  An expression for a general interferometric fringe
pattern is given by \citet{scott} (Fig.~\ref{fringepic}) although for
the purposes of this paper we use a simplified version ignoring
atmospheric effects, and assuming a monochromatic signal
(Eq. \ref{fringeeqn}).

Once a fringe pattern has been recorded, it is useful to obtain the
visibility amplitude and (for space based interferometers) visibility
phase.  If three or more separate apertures have been used, it is
possible to measure the bispectrum (or closure) phase.

\subsubsection{Typical light levels}
The typical light levels required for measurement by an interferometer
are very low, which is why conventional CCDs with higher readout noise
are not acceptable as interferometric detectors.  For example,
observation of a sun-like star with a lossless interferometer at a
distance of 1 kpc, with two telescopes of size 1 m diameter, a single
pixel detector with a sampling rate of 5 kHz, with a bandpass of 30 nm
centered at 700 nm, and a detector with 50 percent quantum efficiency
would produce an average detected signal of about 0.7 photons per
detector pixel per readout from the interference fringe.  It is
therefore essential that an interferometric detector must be able to
detect individual photons if anything other than bright sources are to
be observed.

\subsection{L3CCD multiplication noise}
\label{strategysect}
The on-chip multiplication process in an L3CCD is stochastic and
introduces extra noise into the output signal of the L3CCD before
readout, in addition to the Poisson noise due to the detected photons.
This reduces the signal-to-noise ratio (SNR) of the raw output by up
to $\sqrt{2}$.  By processing the raw measured output \citep{basden}
we can reduce the effect of this stochastic noise at low light levels,
the reduction depending on the processing strategy that we choose.
Increasing the SNR of the L3CCD output signal then allows us to
improve our estimate of interferometric visibilities.

We use the processing strategies suggested by \citet{basden} which
were developed for photometric applications, namely the ``Analogue'',
``PC'' and ``PP'' thresholding strategies which are defined as
follows:
\begin{description}
\item[\textbf{Analogue:}] The L3CCD output is divided by the mean gain.
\item[\textbf{Photon counting, PC:}] If the L3CCD output signal is above some
noise threshold, it is assumed to represent one photon.
\item[\textbf{Poisson probability, PP:}] Thresholds are placed at the positions
where the L3CCD output probability distribution for a given mean input
of $n$ (integer) Poisson photons crosses with the output distribution
for a mean input of $n+1$ Poisson photons.
\end{description}

Additionally we here explore a further thresholding strategy denoted
as ``Uniform'', defined as follows:
\begin{description}
\item[\textbf{Uniform:}] Places the L3CCD output signal into evenly
spaced thresholds, the spacing being dependent on the mean light level
(estimated from the fringe), chosen such that on average, at a given
light level, we will estimate the mean photon input correctly.  This
allows us to make better use of the fringe data.  Further details of
this thresholding strategy are given in appendix~\ref{threshappendix}.
\end{description}

\subsection{Numerical models}
We investigate the effect of using an L3CCD both with and without
processing of the raw output for a range of visibility amplitudes and
light levels, categorizing the performance of L3CCDs as
interferometric detectors, particularly in the estimation of
visibility amplitude and bispectrum (closure) phase.  Any additional
bias introduced to these measurements by the multiplication process
needs to be determined.  In this paper we investigate the effect of
L3CCDs using Monte-Carlo simulation as follows:
\begin{enumerate}
\item Multiple realizations of different fake interferometric fringes
were generated.
\item Poissonization of the fringe signal was simulated.
\item The L3CCD stochastic multiplication process on the
Poissonized signal was simulated.
\item The simulated L3CCD output was processed using techniques known
to reduce the noise introduced by the multiplication process (i.e.\
Uniform, PP, PC).
\item The effect of these processing strategies on fringe visibility
estimation (visibility amplitude and phase, and bispectrum phase) was
investigated using standard methods, and checked to see whether the
estimates were the same as when using a perfect Poissonian detector.
\end{enumerate}

The interference fringe pattern is modelled assuming a two element
interferometer, using the following expression,
\begin{equation}
I_\mathrm{j} = \mu\left[ 1+V\cos\left(2\pi \mathrm{j}m/M +
\phi\right)\right] 
\label{fringeeqn}
\end{equation}
where $I_\mathrm{j}$ is the intensity of the fringe signal in the
$\mathrm{j^{th}}$ pixel, for a fringe with a mean intensity $\mu$ and
visibility amplitude $V$, with phase $\phi$.  The fringe extends over
an array of size $M$ pixels ($M=256$ in our case) and contains $m$
whole wavelengths within this array, so that we are not affected by
aliasing when Fourier transforming.  In our calculations, we ensured
that there were always more than eight samples between adjacent peaks
of the fringe, so that they were well sampled.  Poisson photons with
mean intensity $I_\mathrm{j}$ were then produced from the theoretical
fringe for a given OPD, giving the fringe as it would appear to a
detector before multiplication and readout.  By modelling the
behaviour of the L3CCD multiplication register using a Monte-Carlo
simulation following \citet{basden}, the effect of L3CCDs on
visibility parameter estimation was investigated.

\subsection{Visibility amplitude estimation}
\begin{figure}
\includegraphics[width=8.5cm]{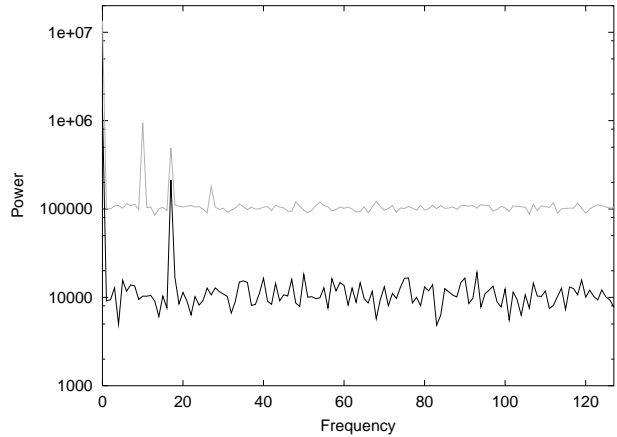}
\caption{Sample powerspectra of interference fringes produced from
L3CCD simulations.  The lower (black) trace is produced by a two
element interferometer with a mean light level of 1 photon per pixel,
and a visibility amplitude of 0.75 from the addition of 20 fringe
powerspectra.  The upper (grey) trace is produced by a three element
interferometer with a mean light level of one photon per pixel, and
visibility amplitudes of 0.75, 0.5 and 0.25 from the addition of 200
fringe powerspectra.  The white noise background can be seen clearly,
and shows that the multiplication process does not alter the slope of
this noise.}
\label{powerspecgraph}
\end{figure}

It is well established \citep{goodman} that the mean power spectrum
(square of the Fourier transform) of an interference fringe averaged
over many realizations will yield a DC peak with height $N^2$ (where
$N$ is the mean number of detected photons in a single fringe), a peak
at a position corresponding to the fringe modulation frequency, and a
white noise background due to Poisson noise and L3CCD multiplication
noise, as shown in Fig.~\ref{powerspecgraph}.  In the case of a two
element interferometer considering only photon noise, an unbiased
estimate of the visibility amplitude \citep{goodman} is obtained using
\begin{equation}
|V|=2\sqrt{\frac{E[D^{(2)}(u)]-b}{E[D^{(2)}(0)]-b}}
\end{equation}
where $E[D^{(2)}]$ is the power spectrum of the fringe measured at the
modulation frequency, $u$, or DC, and $b$ is the noise background level.
L3CCD multiplication noise may introduce a bias to the estimator
depending on the output thresholding strategy and light level.

In some cases, detector imperfections can introduce a slope to the
noise background \citep{perrin}.  However, we have established that
these effects are not relevant when using an L3CCD and so we do not
consider such effects here.

\citet{goodman} give estimates for the number of photon detections that are
required to measure the visibility amplitude of a fringe to a
specified accuracy.  To measure a visibility amplitude $V$ to a SNR of
$\mathcal{R}$ in the photon limited case, we require
\begin{equation}
N = \frac{2\mathcal{R}^2}{V^2}
\label{goodmanstuff}
\end{equation}
total signal detections.  We therefore generate fringes and sum the
power spectra until the total number of Poisson photon detections
allows us to reach the required SNR (in our case 50).  We then use the
summed power-spectrum to estimate the visibility amplitude.  We
investigate visibility amplitudes between $0.02-1$ at mean light
levels between $0.01-20$ photons per pixel, chosen in the range where
\citet{basden} find L3CCD output thresholding has a positive effect on
the SNR.

\subsection{Visibility phase estimation}
The visibility phase is meaningful only for space based
interferometers, since atmospheric perturbations render it meaningless
unless the phase introduced by the atmosphere is known.  It is
estimated from the phase of the Fourier transform at the fringe
modulation peak.  When averaging multiple realizations of the fringe
pattern, we average the complex value of the peak, not just the
phase.  This ensures that we give the correct weighting to each
individual value, dependent on the number of photons detected.

\subsection{Bispectrum phase estimation}
\label{bispecsect}
When light from at least three apertures is combined at different
(non-redundant) OPD modulation frequencies, we can use the detected
signal to retrieve some of the phase information by summing together
the phases, cancelling the relative phase introduced by the
atmosphere.  To model this we use a simplified model of the
generalized fringe pattern given by \citet{scott}.  The fringe power
spectrum will contain three signal peaks and a DC peak as shown in
Fig.~\ref{powerspecgraph} (top trace).  The bispectrum phase is
estimated from the phase of the average complex bispectrum product,
with a bias correction that must be applied at low light levels.  A
suitable correction to apply to the measured bispectrum in the case
when detection is governed solely by photon noise is given by
\citet{wirnitzer}:
\begin{eqnarray}
\label{wirnitzereq}
<Q^{\left(3\right)}\left(u,v\right)> &=& 
E\left[D^{\left(3\right)}\left(u,v\right)\right] + c_1 N \\
&&-c_2 E\left[ | D^{(2)}\left(u\right)|^2 +
| D^{(2)}\left(v\right)|^2 \right.\nonumber \\
&& \left.\;\;\;\;\;+| D^{(2)}\left(-u-v\right)|^2\right]\nonumber
\end{eqnarray}
where $c_1=2$, $c_2=1$, $<Q^{\left(3\right)}\left(u,v\right)>$ is the
unbiased estimate for the bispectrum and
$E\left[D^{\left(3\right)}\left(u,v\right)\right]$ is the expected
triple product of the fringe (the product of the Fourier transform of
the data evaluated at the three signal peaks).  We now evaluate the
phase of $Q^{\left(3\right)}\left(u,v\right)$ instead of the
photon noise biased value of
$D^{\left(3\right)}\left(u,v\right)$, with $u$ and $v$
corresponding to the different fringe frequencies, $D^{(2)}(u)$
representing the fringe power spectrum, and $N$ being the total
photon count.  Eq.~\ref{wirnitzereq} shows that unlike the power
spectrum, the photon bias introduces both  constant and frequency
dependent terms into the bispectrum.

If the form of the noise is not Poissonian, as is the case when using
an L3CCD, a different bias correction may be required.  We are able to
use results from \citet{pehlemann} to determine the bias correction
for the case when we use the raw L3CCD output, obtaining $c_1=6$ and
$c_2=2$.  The details, which involves considering the L3CCD output
probability distribution, are discussed in appendix
\ref{pehlemannappendix}.

If we do not know the detector statistics, it may still be possible to
obtain an unbiased estimation of the bispectrum phase.  To do this, we
used the data to compute $c_1$ and $c_2$ in Eq.~\ref{wirnitzereq}.
This technique relies on Eq.~\ref{wirnitzereq} being the correct model
for the form of the bias correction, and assumes that the power
spectrum can be written as
\begin{equation}
E[D^{(2)}(u)] = c_{PS,0} <J^{(2)}(u)> + c_{PS,1}
\end{equation}
where $c_{PS,0}$ and $c_{PS,1}$ are scalar coefficients, and
$<J^{(2)}(u)>$ is the average power spectrum of the normalized
high-light-level (unbiased) fringe.  

The coefficients $c_1$ and $c_2$ were obtained by evaluating the
bispectrum at frequencies $u,v$ chosen such that other terms
disappear.  We give further details in
appendix~\ref{pehlemannappendix}, and such a scheme was initially
proposed by \citet{pehlemann}.  The form of the power spectrum assumed
by this scheme is not always correct, though since differences will be
small, we attempted to use this scheme to estimate the bispectrum phase.
In most circumstances we find that a correction can be found.

\subsection{Theoretically predicted visibility amplitude bias}
\label{classicvis}
If we use PP or Uniform thresholding strategies on the L3CCD output we
will introduce a systematic bias into our visibility amplitude
estimations at some light levels since we place different weightings
on fringe minima and maxima.  We can get an understanding of how to
treat this by considering a theoretical model of the L3CCD output
probability distribution as given by \citet{basden}:
\begin{equation}
P(x)=\sum_{n=1}^{\infty}\frac{\exp{(-\mu)}
\mu^nx^{n-1}\exp{(-x/g)}}{n!(n-1)!g^n}
\end{equation}
where $g$ is the mean gain, $\mu$ is the mean light level and $x$ is
the number of electrons at the L3CCD output after multiplication.
We can therefore find the estimated light level $I$ in terms of the true
light level $\mu$ for any thresholding strategy by writing
\begin{equation}
I = \sum_{a=1}^\infty a \sum_{x=f_{a-1}}^{f_a} P(x)
\end{equation}
where the $a^\mathrm{th}$ threshold boundary is represented by $f_a$
and $P(x)$ is given above.

The classical definition for visibility is
\begin{equation}
V=\frac{I_\mathrm{max}-I_\mathrm{min}}{I_\mathrm{max}+I_\mathrm{min}},
\label{classicviseq}
\end{equation}
at a given light level
$\mu=\left(I_\mathrm{max}+I_\mathrm{min}\right)/2$ and true visibility
$V$.  The bias in the estimated visibility amplitude can then be
determined, and so corrected, giving us an unbiased estimate for
visibility amplitude.  We expect our visibility estimate
$V_\mathrm{e}$ to be equal to:
\begin{equation}
V_\mathrm{e}(\mu,V) = \frac{
\sum_{a,x,n}
aS(x,n)\left(W_+-W_-\right)
}{
\sum_{a,x,n}
aS(x,n)\left(W_++W_-\right)
}
\label{theoreticalvisbias}
\end{equation}
with
\begin{eqnarray}
S(x,n)&=&\frac{\mu^n x^{n-1}\exp{(-x/g)}}{n!(n-1)!g^n},\nonumber \\
W_\pm &=&\exp(\mp \mu V)(1\pm V)^n \nonumber \\
I_{\mathrm{max}} &=&\sum_{a,x,n} aS(x,n)W_+ \nonumber
\end{eqnarray}
where the summation is over $n=1\rightarrow \infty$,
$x=f_\mathrm{a-1}\rightarrow f_\mathrm{a}$ and finally $a=1\rightarrow
\infty$.  We have used this theoretical model alongside our
Monte-Carlo simulations to verify the predictions we make.

In a similar way, we can compute the visibility amplitude bias when
using a single threshold.  Since we interpret any signal as a single
photon, at a given true light level $\mu$ we measure the mean light
level to be $1-\exp (-\mu)$.  The true light level at the fringe
maxima and minima is given by $I_{\pm}=\mu (1\pm V)$, and so inserting
this into the expression for visibility gives
\begin{equation}
V_{e} = \frac{\exp(-\mu) [\exp(\mu V)-\exp(-\mu V)]}
{2-\exp(-\mu)[\exp(\mu V)+\exp(-\mu V)]}
\label{pcvisbias}
\end{equation}
where the symbols are as before.

\section{Accuracy of fringe parameter estimation}
We separate our analysis into three light level regimes, which allows
us to apply different L3CCD output thresholding strategies as
appropriate for maximum reduction of the noise introduced by the L3CCD
multiplication process.  Our light level regimes are chosen according
to:
\begin{description}
\item[Low light levels:] Much less than one photon per pixel per
readout ($<0.1$ photons per pixel).
\item[Intermediate light levels:] Between about $0.1-20$ photons per
pixel per readout.
\item[High light levels:] More than 20 photons per pixel.
\end{description}

Read noise and gain were found to have little effect on the visibility
estimations, provided the mean gain was significantly greater than the
read noise as found by \citet{basden}.  If this was not the case, then
the L3CCD output signal was dominated by noise, leading to inaccurate
visibility estimation.

We find that all of the L3CCD output thresholding strategies leave the
mean power spectrum of the fringe with a flat background.  The height
of this background is determined by the variance of the signal and in
a Poisson case it is equal to the number of photons detected (which is
also the variance).  When using an L3CCD analogue processing strategy,
we find that the background level is twice the height of that from a
pure photon noise case as expected since the signal variance is double
that of the Poisson case.  Similarly, with a single threshold we find
that the background level is lower than in the pure photon case by an
amount in agreement with the reduction in signal variance.  Multiple
thresholding strategies give a background level above the pure photon
case and below the analogue case, the height depending on light level,
in agreement with the variance of the signal \citep{basden} which is
dependent on the light level.

\subsection{Visibility amplitude estimation at low light levels ($<0.1$ photons per pixel)} 
\begin{figure*}
\includegraphics[width=17cm]{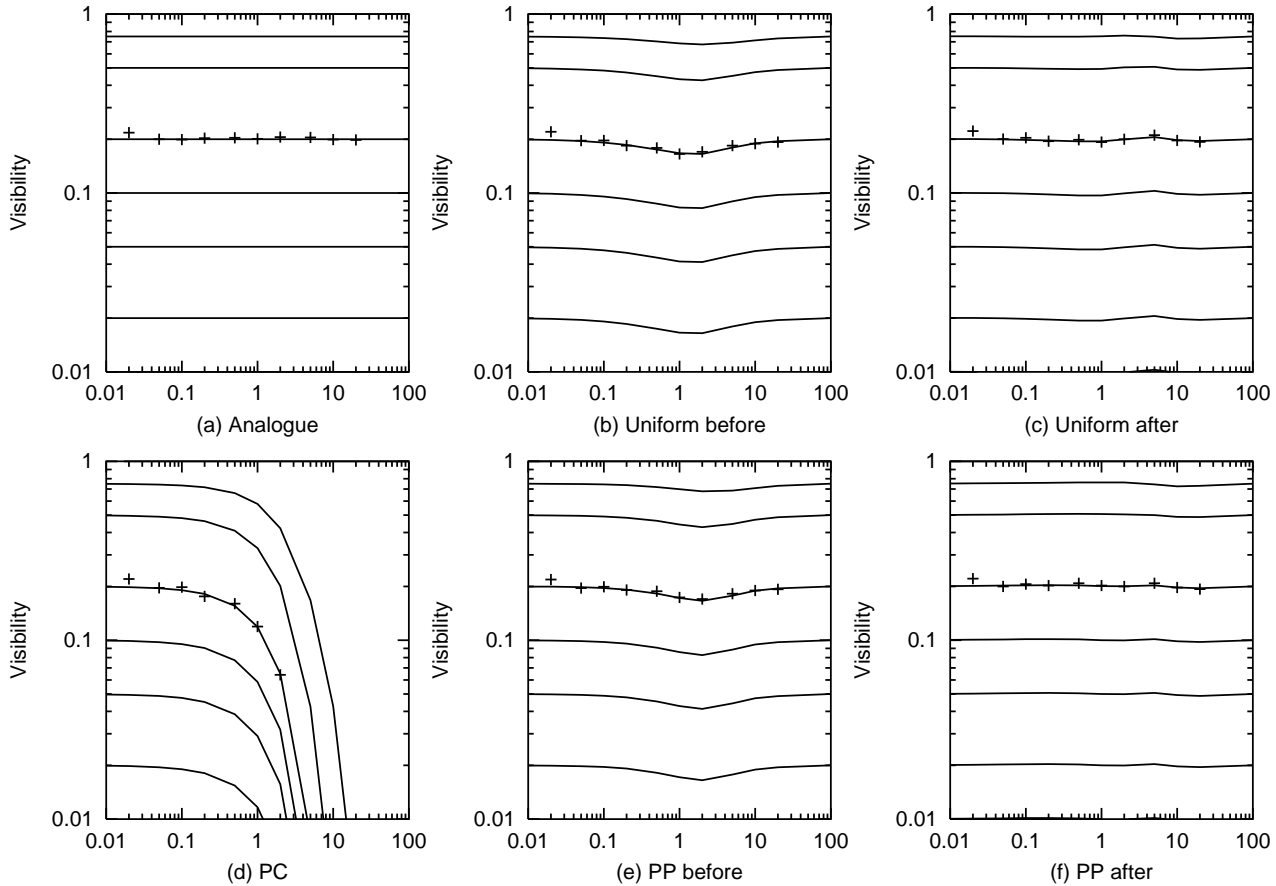}
\caption{Plots showing visibility amplitude estimates when using an
L3CCD as a function of light level (mean photons per pixel), for
visibilities of 1, 0.75, 0.5, 0.2, 0.1, 0.05 and 0.02, with points
representing Monte-Carlo simulation results (SNR 50), and curves being
theoretical.  (a) Results when using an analogue thresholding
strategy.  In this case no bias is evident. (b) and (c) Showing
results when using a uniform thresholding strategy without and with a
correction required to estimate amplitude correctly.  (d) Showing
results using a PC thresholding strategy.  Here we see that this
cannot be used at light levels greater than about $0.1-0.5$ photons
per pixel.  (e) and (f) Showing visibility amplitude estimates when
using a PP thresholding strategy, before and after correction for
visibility amplitude bias respectively.}
\label{vislightgraph}
\end{figure*}

As described by \citet{basden}, if the mean light level is low (much
less than one photon per pixel per readout) then it is most probably
that either one or zero photons will land on a pixel during an
exposure.  At the L3CCD output we will then have either a relatively
large signal corresponding to one photon, or a small signal, just due
to noise from the electronic readout amplifier.  We can therefore use
a PC thresholding strategy treating every signal above a noise level
as representing one photon and operating the L3CCD in photon counting
mode.  We find that the visibility amplitude estimations are unbiased
for all visibilities provided the light level is kept low, as shown in
Fig.~\ref{vislightgraph} (d).

A uniform thresholding strategy will lead to the same results, since
at low light levels it is virtually identical to a PC thresholding
strategy.  Similarly using an analogue strategy (the raw output),
though requiring about twice as many detections to achieve the same
SNR due to a larger excess noise factor \citep{basden}, will also
provide unbiased visibility estimation.  A PP thresholding strategy
also provides an unbiased estimate for visibility.

As the light level increases, using a single threshold (PC strategy)
causes us to lose estimation accuracy, since there is an increased
probability of two or more photons being detected in any given pixel.
These signals will be interpreted as a single photon event, and so the
fringe maxima will be underestimated more than the fringe minima.
Visibility amplitude will be underestimated, and at high light levels
will tend towards zero.  Other thresholding strategies are then more
appropriate.

\subsection{Visibility amplitude estimation at intermediate light levels ($0.1-20$ photons per pixel)}
At intermediate light levels, we cannot use a single threshold
processing strategy as coincidence losses are large.  However, we can
still reduce the excess noise introduced by the multiplication process
by thresholding the L3CCD output signal.

\subsubsection{PP and uniform thresholding}
We can threshold the L3CCD output signal into predetermined thresholds
which are chosen to minimize the excess noise factor introduced by the
stochastic multiplication process, for example using the PP
thresholding strategy \citep{basden}.  Alternatively, we can use the
raw data from a fringe to estimate the mean light level.  We can then
use this along with our knowledge of the mean gain to improve our
input prediction, estimating the most likely value of the photon input
by choosing appropriate thresholds taking into account the mean light
level (appendix~\ref{threshappendix}).  For any given pixel, the
number of photon events may be estimated incorrectly, though on
average we will estimate correctly.  At low signal levels, we will
estimate correctly most of the time, minimizing the dispersion
introduced by the multiplication process.  As the light level
increases, we will lose accuracy progressively \citep{basden} until
there is no advantage in thresholding over using the raw L3CCD output.

Using the PP or Uniform thresholding strategies at light levels
between about $0.1-10$ photons per pixel will introduce a small bias
to the estimated visibility amplitude.  We find that this bias is
greatest at light levels of about two photons per pixel, and that here
the estimated visibility is just over 80 percent of the true
visibility.  This is independent of the mean gain, provided the gain
is well above the readout noise.  An explanation for this bias is
given in appendix~\ref{visbiasappendix}, and we find that our
Monte-Carlo simulations agree with the theoretically prediced bias
(section \ref{classicvis}).  We find that a simple bias correction can
be applied according to
\begin{equation}
V_\mathrm{cor}\approx
\frac{V_\mathrm{e}}{V_\mathrm{e}^{5/2}\left(1-h(\mu)\right)+h(\mu)}
\end{equation}
where $h(\mu) = 1-\mu^{3/4}\exp(-\mu/3)/5$ for a mean light level
$\mu$ photons per pixel, with $V_\mathrm{e}$ representing the biased
visibility estimated from the thresholded data, and $V_\mathrm{cor}$
being the corrected visibility amplitude.  It is important to note
that this is only an approximation generated by fitting to the data,
though it allows the visibility amplitude to be estimated to about two
percent accuracy, i.e.\ the visibility will be estimated to within
$(1\pm0.02)V_\mathrm{true}$.  A more accurate bias correction can be
estimated by considering the L3CCD output probability distribution,
though this requires analytical treatment to compute it.

Monte-Carlo simulation results before the application of the simple
bias correction, along with the variation of
Eq.~\ref{theoreticalvisbias} and Eq.~\ref{pcvisbias} with $\mu$ and
$V$ are shown in Fig.~\ref{vislightgraph} (b), (d) and (e) for the
uniform, PC and PP thresholding strategies.  Fig.~\ref{vislightgraph}
(c) and (f) show the visibility amplitude estimates after the simple
bias correction has been applied.  We find that the visibility bias is
very similar for both the uniform and PP thresholding strategies.

\subsubsection{Raw output}
Using the analogue processing strategy provides an unbiased estimate
for visibility, since we are not weighting the data in any way.  This
would seem to be a big advantage, except that the SNR can be reduced
by a factor of up to $\sqrt{2}$ when compared with other thresholding
strategies.  We therefore recommend the use of this processing
strategy at these light levels in all cases where there is ample SNR.
Visibility amplitude reproduction is shown in Fig.~\ref{vislightgraph}
(a).

\subsection{Visibility amplitude estimation at high light levels}
At high light levels, there is no real advantage in thresholding the
L3CCD output as this does not improve the SNR, and so an analogue
processing strategy should be used, treating the L3CCD output as that
from a conventional CCD.  Visibilities (amplitude and phase) will be
estimated with no additional bias.

\subsection{Phase}
The visibility phase estimate from a fringe produced by a two element
interferometer is unbiased in all of the previously mentioned readout modes
(section~\ref{strategysect}) at any light level.  It is only fringe peak
height, and not position of the fringe that is altered by thresholding
the raw L3CCD output, and so the L3CCD has no effect on the visibility
phase estimation.

\begin{figure*}
\includegraphics[width=17cm]{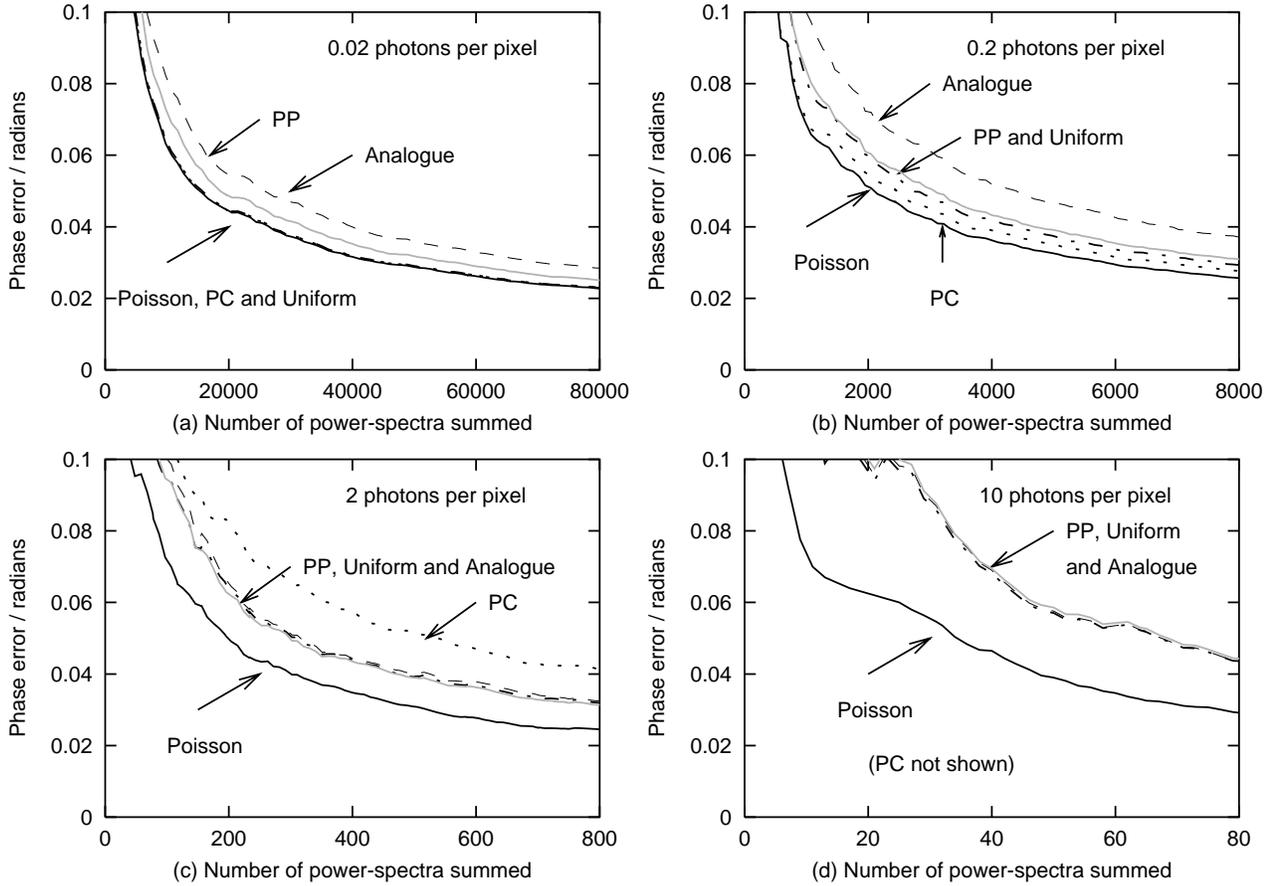}
\caption{Visibility phase error plotted against the number of fringe
patterns recorded and processed when using an L3CCD as a detector.
The solid black curves are an example of the phase error that may be
obtained when using a perfect detector and closely matches
Eq.~\ref{goodmanstuff}.  The dashed curves show the phase error with an
L3CCD and an analogue thresholding strategy, the dotted curves for a
PC thresholding strategy, the dot-dashed curve for a uniform
thresholding strategy and the grey curves for a PP thresholding
strategy.  All plots are for a visibility amplitude of 0.1, and the
mean light level is (a) 0.02 photons per pixel, (b) 0.2 photons per
pixel, (c) 2 photon per pixel and (d) 10 photons per pixel.  We can
see that the analogue thresholding strategy generally requires more
photons to reduce the phase error to a given value.}
\label{phaseerrorgraph}
\end{figure*}

The spread in a phase measurement estimated from many fringes is given
by \citet{buscher} as:
\begin{equation}
\alpha = \frac{\sqrt{<\sin^2 \epsilon>}}{\sqrt{n}<\cos \epsilon>}
\end{equation}
where $\epsilon$ is the deviation of the phase estimate from the true
phase in each fringe pattern, and $n$ is the number of phase measurements
(fringes).  We find that $\alpha$ generated by our simulations agrees
well with the theoretical SNR $\alpha^{-1} = NV^2/2$ \citep{goodman}, when
the excess noise factor introduced by the L3CCD multiplication process
after thresholding is taken into account.  An example from our
simulations is shown in Fig.~\ref{phaseerrorgraph}.

\subsection{Bispectrum phase}
\begin{figure*}
\includegraphics[width=17cm]{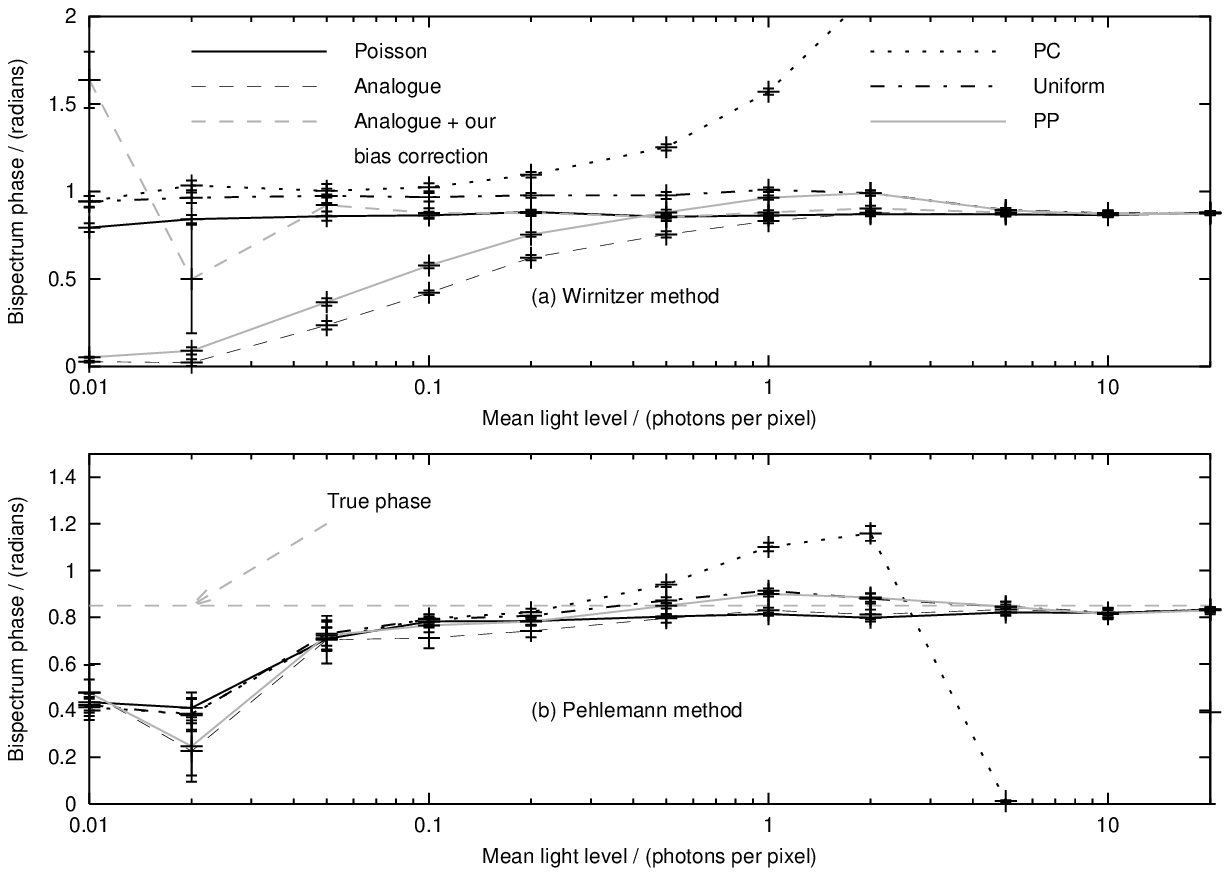}
\caption{Estimated bispectrum phase when using an L3CCD comparing the
performance of different threshold processing strategies (represented
by different curves) and estimation techniques (separate plots).  (a)
We see that using the standard Wirnitzer correction, we are unable to
estimate the bispectrum phase at low light levels when using analogue or
PP thresholding strategies.  However, by using our bias correction, we
are able to use the analogue strategy to estimate the phase.  We also
see that the PC and Uniform thresholding strategies tend to
overestimate phase slightly at low light levels.  (b) Using the
bispectrum phase estimation technique developed by \citet{pehlemann}
we are unable to estimate phase accurately at light levels less than
about 0.1 photons per pixel.  The Poisson curve represents estimation
when using a perfect detector, and the true bispectrum phase is 0.85 radians.}
\label{bispecphasegraph}
\end{figure*}

Our simulations show that when using an L3CCD we are able to estimate
the bispectrum phase accurately at mean light levels down to about
five photons per pixel when using the standard photon noise bias
correction (Eq.~\ref{wirnitzereq}).  At lower light levels, the
bispectrum phase estimate appears to be biased
(Fig. \ref{bispecphasegraph}).  If an analogue or PP thresholding
strategy is used with the standard photon bias correction, we find
that the estimate of phase tends towards zero as light decreases, even
if photometric corrections are applied when using the PP strategy.  

In the analogue case, we are able to estimate phase correctly by using
the bias correction that we adapted from \citet{pehlemann}
(Eq.~\ref{wirnitzereq} with $c_1=6$ and $c_2=2$).  
At very low light levels, a large number of fringe patterns need to be
recorded to reduce the SNR to acceptable levels.

Using a uniform thresholding strategy and the standard
\citet{wirnitzer} bias correction, we find that the phase is
overestimated slightly at low light levels as shown in
Fig.~\ref{bispecphasegraph} (a).  A PC thresholding strategy used with
the \citet{wirnitzer} bias correction is found to overestimate the
phase slightly up to light levels of about 0.2 photons per pixel,
above which the estimate tends quickly towards $\pm\pi$ as coincidence
losses become large.

The failure of Eq.~\ref{wirnitzereq} when estimating the phase is due
to the L3CCD output probability distribution no longer being
Poissonian, which is required for the correction to work correctly.
However, by calculating appropriate values for the coefficients in
Eq.~\ref{wirnitzereq} suitable for an analogue processing strategy, we
are able to obtain an unbiased estimation of bispectrum phase when
using the this processing strategy down to very low light levels.  

A similar correction for other L3CCD output processing strategies has
not been determined due to the complicated nature of the probability
distributions.  However, by using a different approach to bispectrum
estimation, using the data itself to calculate the bias terms
(\citet{pehlemann}, section \ref{bispecsect} and appendix
\ref{pehlemannappendix}), we find that we are able to estimate
bispectrum phase accurately to light levels down to about 0.1 photons
per pixel when using the PP or uniform thresholding strategies.  This
corresponds to the light levels at which L3CCDs are most likely to be
used.  Below this, errors in the individual bias terms, and invalid
assumptions about the form of the power spectrum reduce the phase
estimate below the true value, as shown in Fig.~\ref{bispecphasegraph}
(b).

We recommend the use of the analogue processing strategy and our
adapted bias correction as this gives an unbiased estimate of
bispectrum phase.  Alternatively, if the SNR is low or at very low
light levels, we recommend the use of the Uniform thresholding
strategy since this is able to reduce the effect of stochastic
multiplication noise.  The Uniform thresholding strategy will however
lead to a slight overestimation of the bispectrum phase by between
about $10-20$ percent at the lowest light levels.

\subsection{Uncertainties in gain}
A knowledge of the mean gain is required when using either the PP or
uniform thresholding strategies.  If this is not known precisely,
threshold boundaries may be placed wrongly.  Fortunately, due to the
nature of visibility, being composed of both a difference and a ratio
(Eq.~\ref{classicviseq}), the effect on the amplitude estimation will
be much less than the error in gain.

We find that when using a PP thresholding strategy, the uncertainty in
estimated visibility amplitude is inversely proportional to the
uncertainty in the gain estimate.  If we overestimate the gain by 20
percent, then in the worst case (for low visibilities ($<0.1$) and
light levels around two photons per pixel), we will underestimate the
visibility amplitude by at most four percent for a PP thresholding
strategy, and five percent for a uniform thresholding strategy.  For
example, if the true mean gain was 1000, but we had estimated it to be
1200, and are trying to measure a true visibility amplitude of 0.05 at
a mean light level of 2 photons per pixel, we would estimate the
visibility amplitude wrongly by about five percent, i.e. we would
estimate the visibility amplitude to be 0.0475.

For visibilities above about 0.1, and mean light levels fainter or
brighter than two photons per pixel, the error in visibility is
reduced.  For PC and analogue thresholding strategies, uncertainties
in the mean gain have no effect on visibility estimation.

Typical uncertainties in the gain are of order one percent
\citep{mackay}, and so in general there will be negligible uncertainty
in the visibility estimate due to this uncertainty.  Visibility phase
estimates are unaffected by uncertainties in the mean gain.

\section{Conclusions}
We have investigated the application of L3CCDs to interferometric
fringe detection, and the way in which this affects visibility
parameter estimation.  In doing this, we have introduced an additional
(to \citet{basden}) L3CCD output thresholding strategy to enable
bispectrum phase estimation at low light levels.

In summary, we find that:
\begin{enumerate}
\item L3CCDs can be used for visibility estimation at any light level,
though a small bias correction should be applied if a multiple
threshold processing strategy is used on the L3CCD output.  
\item L3CCD visibility amplitude prediction at high (greater than
about ten photons per pixel) and very low (less than 0.1 photons per
pixel) photon rates is unbiased for all thresholding strategies.
\item The use of multiple thresholding strategies will introduce a
small bias for visibility amplitude at light levels between 0.1-20
photons per pixel.
\item Since the mean gain is typically known to one percent accuracy,
the error in visibility amplitude and phase estimation due to this
uncertainty will be minimal.
\item At light levels greater than about five photons per pixel, the
high-light-level bispectrum phase estimate can be obtained when using
L3CCDs.  At lower light levels down to about 0.1 photons per pixel, a
bias term in the bispectrum phase becomes non-negligible, though this
can be corrected by using the data to calculate the bias terms.  Below
this, a bias correction is still available when the raw L3CCD output
is used, giving an unbiased estimate of the bispectrum phase.
\item Using the raw L3CCD output with our adapted bispectrum bias
correction, or a uniform thresholding strategy with the standard bias
correction allows best bispectrum phase estimation.
\end{enumerate}
Our recommendation is that visibility amplitude and phase estimation
at low light levels (less than 0.1 photons per pixel per readout)
should be carried out using a single threshold processing strategy on
the L3CCD output.  At higher light levels, an analogue processing
strategy should be used if there is ample signal-to-noise.  If the SNR
is low, using a multiple thresholding strategy with bias correction
will help to improve the visibility estimate.  Bispectrum phase
estimation should be carried out using either the unthresholded L3CCD
output or a uniform multiple thresholding strategy with the
corresponding bias correction.

Since L3CCDs provide the most accurate input estimation at low light
levels using a single threshold, we recommend that, if possible, they
are always used in this regime, increasing the frame rate if necessary
to keep the number of photons per pixel low ($<0.1$).  We are
currently developing a new controller for L3CCDs, allowing pixel rates
of up to 30~MHz, which will allow signal levels to be kept low in most
interferometric applications.

\appendix
\section{L3CCD uniform thresholding strategy}
\label{threshappendix}
\citet{basden} provide information about thresholding strategies which
can be used to increase the SNR in a signal from an L3CCD.  For the
purposes of this paper, we here introduce a new thresholding strategy
which uses a knowledge of the mean light level to further improve the
L3CCD signal SNR.  The thresholds are placed with uniform separations,
the separation being dependent on light level, chosen so that the mean
photon flux is estimated correctly.

\subsection{Determining uniform threshold separations}
When using L3CCDs for interferometric signal detection, we are able to
obtain additional information which will allow us to develop a more
accurate thresholding strategy.  Since we make measurements of the
interference signal over a fringe, we are able to estimate the mean
light level within this fringe.  We can then use this knowledge along
with our knowledge of the mean gain to set the thresholds.  Such a
thresholding strategy is therefore dependent on the light level and so
is only useful when an estimate of this can be obtained reliably in
advance.  We use a uniform threshold step size (with threshold
boundaries being uniformly spaced) sized such that the mean
thresholded output signal will be equal to the mean input (though we
will not always threshold correctly for each event).  We then
interpret an L3CCD output signal falling between the $n^\mathrm{th}$
and $n+1^\mathrm{th}$ threshold as representing $n$ photons.

The probability of obtaining an output $x$ electrons from the
multiplication register of an L3CCD with a mean (Poisson distributed)
input, $\mu$, is given by \citep{basden}:
\begin{equation}
P(x)=\sum_{n=1}^{\infty}\frac{\exp{(-\mu)}
\mu^nx^{n-1}\exp{(-x/g)}}{n!(n-1)!g^n}
\label{probdisteqn}
\end{equation}
where $g$ is the mean gain and $\mu$ is the mean light level.

Thresholding this signal can improve the noise statistics.  We can
represent this by
\begin{equation}
P(a) = \sum_{x=f_{a-1}}^{x=f_a} P(x)
\end{equation}
where $P(a)$ is the probability that a signal is put into the
$a^{th}$ threshold, which has boundaries determined by $f_a$.

The expected output is then
\begin{equation}
\overline{a} = \sum_{a=1}^{\infty} a P(a).
\end{equation}

If we require a thresholding scheme which is able to predict photon
flux accurately without any additional corrections, we require the
threshold boundaries to be dependent on light level.  We equate the
above equation to the mean light level, $\mu$, using the threshold
boundaries as our degree of freedom.  If we wish to have uniformly
spaced boundaries ($f_a=aT$ for threshold step size $T$), we can
express this as
\begin{eqnarray}
\mu &=& \frac{\exp{\mu}}{g} \sum_{n=1}^{\infty}\left\{
\frac{\mu^n}{n!(n-1)!} \right.\nonumber \\
& & \times\left.  \sum_{a=1}^{\infty} \left[ a
\sum_{x=(a-1)T}^{x=aT} \left(\frac{x}{g}\right)^{n-1} \exp (-x/g)
\right] \right\}
\end{eqnarray}
which should be satisfied by solving $T$ for a given $\mu$.

A good approximation for the theoretical threshold step size at light
levels greater than about 0.1 photons per pixel, $T$, is
given by
\begin{equation}
T \approx g \left(1+\frac{1}{2 \mu^{3/4}}\right)
= g \left(1+\frac{g^{3/4}}{2 x^{3/4}}\right)
\label{thresholdsize}
\end{equation}
for gain $g$, and mean light level $\mu=x/g$ measured in expected
photons per pixel per readout estimated from an L3CCD output of $x$
electrons.  At light levels less than about 0.1 photons per pixel it
results in a larger threshold size than the ideal, though this is not
problematic since at these light levels we are effectively using a
large single threshold size, and input estimations are almost
identical to those using a PC thresholding strategy.

Using this thresholding strategy, the excess noise factor as defined
by \citet{basden} is unity at low light levels, so the SNR scales as
$\sqrt{n}$.  This is because the step sizes here are large, and so we
make correct predictions most of the time.  As the light level
increases, the excess noise factor tends towards $\sqrt{2}$, the
SNR scaling as $\sqrt{n/2}$ (as with the analogue
case), halving the effective quantum efficiency of the L3CCD.

For cases where the mean light level can be estimated in advance of
thresholding, this processing strategy leads to an improvement over
the PC threshold strategy, being applicable at any light level, and
results in an increased SNR over the analogue and PP strategies at low
light levels, where the improvement is most welcome.  It also has the
advantage that no additional correction should be made after the event
to preserve flux, since it is defined to estimate flux correctly.

\subsection{Errors in threshold step size}
If the light level or gain are not known precisely, errors may be made
when choosing the threshold step size.  If the threshold step size is
too small, we overestimate the photon count, while if it is too large,
we underestimate.

It is straightforward to show that the error in threshold step size
depends on the error in gain and light level as:
\begin{eqnarray}
\left(\frac{\Delta T}{T}\right)^2 &=& \left(\frac{\Delta g}{g}\right)^2
+ \frac{9}{16} \left(\frac{\Delta \mu}{\mu}\right)^2 \nonumber \\
&=& \frac{25}{16}\left(\frac{\Delta g}{g}\right)^2 +
\frac{18}{16}\frac{1}{m^3 n}
\end{eqnarray}
where $\Delta$ represents the uncertainty in a measurement, and the
second equality is valid if the mean light level $\mu$ is calculated
from $m$ averaged raw L3CCD outputs, each containing a mean of $n$
photons with a SNR of $\sqrt{n/2}$ due to the
Poisson and multiplication processes.  We can usually expect to know
the mean gain to about one percent \citep{mackay}, and so this error
is the limiting factor when $m^3 n \ga 10000$, for example
$n=0.0005$ when $m=256$.  We can therefore usually expect to be able
to choose our threshold step size with an accuracy equal to that in
the mean gain.

At high light levels, the fractional error in prediction is
proportional to the fractional error in threshold step size.  At low
light levels, there is less of a dependence since the thresholds are
widely spaced, and signals usually fall within the first threshold.

\section{Multiple threshold visibility amplitude bias}
\label{visbiasappendix}
If we use multiple thresholding on the L3CCD output we will introduce
a systematic bias into our visibility amplitude estimations, though
we can counter for this effect since we can determine the bias.  The
reason for this bias is now discussed.

\subsection{PP thresholding}
When we use a PP thresholding strategy, to estimate the flux
correctly we must apply a correction according to \citep{basden}
\begin{equation}
I_\mathrm{e}/I_\mathrm{r}\approx
\left[1+0.7\exp(-I_\mathrm{e}/3)\right].
\label{ppfluxbias}
\end{equation}
with subscripts e and r corresponding to estimated and real values
respectively.  Since this correction depends on the light level, we
are therefore biasing our visibility amplitude estimation since fringe
minima will have a different weighting (requiring a different
correction) than fringe maxima.  Our estimate of visibility amplitude
will therefore be biased.

We are able to verify this by considering the definition of visibility
amplitude, $V=\frac{I_{+}-I_{-}}{I_{+}+I_{-}}$ where $I_{\pm}$ is the
intensity at the maximum and minimum of the fringe.  Since $1\leq
I_\mathrm{e}/I_\mathrm{r} \leq 1.7$ \citep{basden}, we can
further approximate Eq.~\ref{ppfluxbias} to
\begin{equation}
I_\mathrm{e}\approx
I_\mathrm{r}\left[1+0.7\exp(-I_\mathrm{r}/3)\right].
\end{equation}
Therefore, using our estimated flux with $I_\mathrm{r\pm} = \mu(1\pm
V_\mathrm{r})$ for mean light level $\mu$, and inserting into the
definition of visibility amplitude, we will obtain an estimate
($V_\mathrm{e}$) of the true visibility {$V_\mathrm{r}$) equal to:
\begin{equation}
V_\mathrm{e} \approx \frac{2V_\mathrm{r}%
+0.7\left[f_{+}
\exp\left(-\mu f_{+}/3\right) -%
f_{-} \exp\left(-\mu%
f_{-}/3\right) \right] }%
{2+0.7\left[ f_{+}
\exp\left(-\mu f_{+}/3\right) +%
f_{-}
\exp\left(-\mu f_{-}/3\right)\right]}
\end{equation}
where $f_\pm = 1\pm V_\mathrm{r}$.  We find that this agrees very
closely with our calculation using the probability distributions,
Eq.~\ref{theoreticalvisbias}.

\subsection{Uniform thresholding}
If we use a uniform thresholding strategy, then since we are able to
measure the mean intensity ($\mu$ photons per pixel) of an
interference fringe we use this information to set the threshold step
size.  However, for visibility V, the fringe intensity will actually
vary from a maximum of $\mu(1+V)$ to a minimum of $\mu(1-V)$.  At
these points our threshold step sizes will be wrong.  At a minimum of
intensity, the threshold step size chosen from the mean light level
will be too small leading to an overestimation of signal, while at
intensity maxima the threshold step size will be too large resulting
in an underestimation.  The combined effect is to underestimate the
visibility.

\subsection{Low and high light levels}
Visibility amplitude bias correction is not be required for mean light
levels below about 0.1 photons per pixel, and above about 10 photons
per pixel.  At the lowest light levels, we will generally only be
using the first one or two thresholds, and every photon event will
have the same chance of falling into these.  At higher light levels,
we find that the bias in our visibility amplitude estimation will tend
towards zero, leading to correctly estimated visibility amplitude.

\section{Bispectrum phase bias estimation}
\label{pehlemannappendix}
\subsection{Unbiased bispectrum phase estimation: Method I}
\citet{pehlemann} have studied photon bias compensation of both power
spectrum and bispectrum estimation, and derive a method for correcting
this.  They assume that an image is made from individual photons, each
of which may have a different peak intensity, $a_j$.  We can relate
this to the L3CCD output, using the L3CCD output probability
distribution.  The total expectation value $E[D^{(2)}(u)]$ of the
power spectrum of a data set is given by
\begin{equation}
E[D^{(2)}(u)] = c_{PS,0} <J^{(2)}(u)> + c_{PS,1}
\label{pehlpowspec}
\end{equation}
where $c_{PS,0} = \overline{N}^2 \overline{a}^2$ and $c_{PS,1} =
\overline{N} \overline{a^2}$ are scalar coefficients, and
$<J^{(2)}(u)>$ is the average power spectrum of the normalized
high-light-level (unbiased) fringe.  We find that this expression can
be applied to the L3CCD raw output since it has this form.  However,
when using other L3CCD thresholding strategies, this expression is not
strictly true, though we do indeed get a flat noise background,
$c_{PS,1}$.

The total expectation for the bispectrum, $E[D^{(3)}(u,v)]$ is given
by \citet{pehlemann} as
\begin{eqnarray}
E[D^{(3)}(u,v)] & =& c_{BS,0} <J^{(3)}(u,v)> + c_{BS,2}\nonumber \\
& + & c_{BS,1} \left( <J^{(2)}(u)> + <J^{(2)}(v)>\right. \nonumber \\
&& +  \left.<J^{(2)}(-u-v)> \right)
\label{pehlbispec}
\end{eqnarray}
where $c_{BS,0} = \overline{N}^3\overline{a}^3$, $c_{BS,1} =
\overline{N}^2 \overline{a} \overline{a^2}$, $c_{BS,2} = \overline{N}
\overline{a^3}$ are scalar coefficients and $<J^{(3)}(u,v)>$ is the
desired average bispectrum of the normalized high-light-level
(unbiased) fringe.

Since we know that the probability distribution for the L3CCD output
given a single photon input \citep{basden} is $P(x) =
g^{-1}\exp(-x/g)$, we can calculate the coefficients as:
\begin{eqnarray}
c_{PS,0}&=&\overline{N}^2 g^2\nonumber \\
c_{PS,1}&=&2\overline{N} g^2\nonumber \\
c_{BS,0}&=&\overline{N}^3 g^3\nonumber \\
c_{BS,1}&=&2\overline{N}^2 g^3\nonumber \\
c_{BS,2}&=&6\overline{N} g^3\nonumber
\end{eqnarray}
where $g$ is the mean gain.  When using the analogue thresholding
strategy, we simply divide the L3CCD output by $g$.  This allows us to
write
\begin{eqnarray}
\overline{N}^3<J^{(3)}(u,v)> &=& E[D^{(3)}(u,v)] +
6\overline{N} \nonumber \\
&-& 2\left(E[D^{(2)}(u)]+E[D^{(2)}(v)]\right.\nonumber \\
&&\left.+E[D^{(2)}(-u-v)]\right)
\end{eqnarray}
from which we can obtain an unbiased estimate of the bispectrum phase.

\subsection{Unbiased bispectrum phase estimation: Method II}
If we are not easily able to calculate the probability distribution
for $a$, we can estimate the required correction terms using our data.
To do this, we evaluate the bispectrum at many frequencies chosen so
that we can obtain the required coefficients one at a time.

\begin{enumerate}
\item By choosing two frequencies, $f_1$ and $f_2$ such that $f_1, f_2
\neq u, v, u+v$ where $u$, $v$ and $u+v$ are the frequencies at which
the peaks appear in the power spectrum, we can see that
$E[D^{(3)}(f_1,f_2)] = c_{BS,2}$, and so obtain an estimate for $c_{BS,2}$.

\item Provided $u\neq v$ and $u\neq 2v$ we evaluate the bispectrum at
$f_1=f_2=u,v,u+v$.  At these frequencies, $<J^{(3)}(f_1,f_2)>=0$ and so
the bispectrum here evaluates to $c_{BS,2}+2 c_{BS,1}<J^{(2)}(f_1)>
+c_{BS,1}<J^{(2)}(2f_1)>$.

\item Using our power spectrum (Eq.~\ref{pehlpowspec}) we can estimate
$c_{PS,0}<J^{(2)}(u)>=E[D^{(2)}(u)]-c_{PS,1}$, obtaining $c_{PS,1}$
from the flat noise background.

\item We are now able to estimate $c_{BS,1}/c_{PS,0}$ from the
previous two steps:
\begin{equation}
\frac{c_{BS,1}}{c_{PS,0}} = 
\frac{E[D^{(3)}(u,u)]-c_{BS,2}}{2E[D^{(2)}(u)]-3c_{PS,1}} 
\end{equation}

\item The unbiased bispectrum phase can then be obtained at $f_1=u$
and $f_2=v$, since this is independent of $c_{BS,0}$.  
\end{enumerate}

\bsp
\label{lastpage}

\end{document}